\documentclass[12pt]{article}

\newcommand{\beq}{\begin{equation}}
\newcommand{\eeq}{\end{equation}}
\newcommand{\beqa}{\begin{eqnarray}}
\newcommand{\eeqa}{\end{eqnarray}}
\def\sepand{\rule{14cm}{0pt}\and}

\begin{document}

\title{White dwarf cooling and large extra dimensions}

\vspace{0.6cm}

\author{{\sc Marek Biesiada}\\
\sepand
{\sc Beata Malec}\\
{\sl Department of Astrophysics and Cosmology,}\\
{\sl University of Silesia,}\\
{\sl Uniwersytecka 7, 40-007 Katowice, Poland}\\
{mb@imp.sosnowiec.pl;  beata@server.phys.us.edu.pl}\\
}
\date{}
\maketitle
\vfill

\begin{abstract}

Theories of fundamental interactions with large extra dimensions have recently become very popular.
Astrophysical bounds from the Sun, red-giants and SN1987a have already been derived by other
authors for the theory proposed by Arkani-Hamed, Dimopoulos and Dvali. 
In this paper we consider G117-B15A pulsating white dwarf (ZZ Ceti star) for which the secular rate 
at which the period of its fundamental mode increases has been accurately measured 
and claimed that this mode of G117-B15A is perhaps the 
most stable oscillation ever recorded in the optical band.   
Because an additional channel of energy loss (Kaluza-Klein gravitons) would speed up the cooling rate, one is able 
to use the aforementioned stability to derive a bound on theories with large extra dimensions. 
Within the framework of the theory with large extra 
dimensions proposed by Arkani-Hamed, Dimopoulos and Dvali 
we find the lower bound on string comapctification scale  
$M_s > 14.3 \; TeV/c^2$ which is more stringent than solar or red-giant bounds.

 \end{abstract}

\section{Introduction}

The interest in physical theories with
extra spatial dimensions has recently experienced considerable revival.
Multidimensional theories with compact
spatial dimensions having inverse radius at the order of the GUT scale were
investigated intensively in the 80-ties \cite{Green87}.
Later developments in string theory initially due to Anoniadis \cite{Antoniadis} later supported by 
\cite{Witten} suggested that it could be possible to
have a much lower string or compactification scale.
In particular, it is has been conjectured \cite{Dvali} that compactification scale
could be at the order of a TeV, corresponding to a weak-scale string theory.
Such a low
string scale is attractive from experimental perspective since
the string state spectrum would now become accessible.
In this class of theories, gravity
is essentially $n+4$ dimensional whereas all other physical fields are confined to
4-dimensional brane.
The relation between the Planck mass in 4 dimensions ($M_{Pl} = 1.2\;10^{19}\;GeV/c^2$)
and the string mass scale in $4+n$ dimensions $M_s$ and the radius $R$ of extradimensional
space reads:
\begin{equation} \label{R}
R^n = \left( \frac{\hbar}{c} \right)^n \frac{M_{Pl}^2}{M_s^{n+2}\Omega_n}
\end{equation}
where $\Omega_n$ is the volume of the unit n-sphere. Present laboratory limits \cite{LEP}
give $1 TeV$ as lower bound on $M_s$. Assuming that $M_s$ is of order of $TeV$ (sustaining
the hopes of experimental verification of multidimensionality of the world) one can
immediately rule out $n=1$ because in that case one would expect modified gravity at
distances $R \approx 10^{15}\;cm$ which is not observed.

Last century witnessed a successful application of standard physics in elucidating
the properties of celestial bodies which consequently can be used as a source of -
sometimes strong - bounds.
The idea that astrophysical considerations can constrain ``exotic'' physics is not a new one.
For example it has been implemented to constrain the axion mass \cite{Raffelt}.
The main idea here is that if ``exotic'' physics to be tested predicts the existence of weakly
interacting particles which can be produced in stellar interiors such weakly
interacting particles would serve as an additional source of energy loss in many ways
influencing the
course of stellar evolution. In the context of multidimensional physics
excitation of Kaluza-Klein gravitons
play the role of stellar extra coolant.

There are three main sources of astrophysical bounds invoked in this context. First is
the Sun
which is a hydrogen burning main sequence star with radiative interior. Back reaction in
response to increased energy loss results in rising the internal temperature and shrinking
of the radiative interior (for details see \cite{Raffelt}).
Because enhanced temperature increases
the rates of nuclear reactions this would exacerbate solar neutrino problem. Moreover
helioseismology provides an accurate estimate of the internal profile of isothermal sound speed
squared \cite{helioseismology}. The most accurate value of this quantity, which is proportional to
temperature, refers to the radius $r = 0.2\;R_{\odot}$. This provides an effective mean to
constrain exotic sources of energy loss.

The second source of bounds comes from red giants - stellar evolutionary phase after the
hydrogen has been exhausted and
which lasts until the helium ignition takes place during the so called helium flash.
Additional cooling would results in one of the following effects. First of all
 helium flash would not occur
if cooling was too effective. If additional cooling is less effective (so that helium
flash eventually takes place) the red giant tip on
the Hertzsprung-Russel diagram would be located higher above the Horizontal Branch and the
time spent on a Horizontal Branch would be shortened. All these effects could be tested
on HR diagrams for globular clusters \cite{Raffelt}.

The last and the most effective of traditional sources of bounds comes from SN1987a -
more specifically from the observed duration of neutrino pulse. The pulse would be
shortened
had the nascent neutron star cooled down more rapidly than standard theory predicted.
These considerations have been applied, in the context of large extra-dimensions in
works of Barger et al. \cite{Barger}, Cullen and Perelstein \cite{Cullen}
 and Cassisi et al \cite{Cassisi} (see also \cite{Savage}). 

In  this paper we consider another class of astrophysical objects for which the cooling
rate is known from observations, namely the pulsating white dwarfs.
We will focus our attention on ZZ Ceti star G117-B15A
which 
was considered previously as a tool of testing fundamental physics (in the context of axion emission) 
\cite{Isern, Raffelt}.

\section{White dwarf cooling in the presence of large extra dimensions}

White dwarfs such like G117-B15A are final stages of evolution of stars with
masses smaller than $8\;M_{\odot}.$  
They have inner degenerate core composed of carbon and oxygen which is an internal energy
reservoir and a thin non-degenerate envelope made up of He and H shells. Having exhausted
their nuclear fuel these stars can only contract and cool down. Some of them e.g. ZZ Ceti
stars are pulsating and their pulsation period increases as the star cools down
\cite{Baglin69, Winget83}.
The rate of period increase ${\dot P}/P$ is proportional to temperature decrease rate
${\dot T}/T$. 

At an energy scale much lower than the string scale, one can construct an effective
theory of KK gravitons
interacting with the standard model fields \cite{Giudice}. Barger et al. \cite{Barger}
have calculated Kaluza-Klein graviton emissivities in five processes interesting from
astrophysical perspective: photon-photon annihilation, electron - positron annihilation,
Gravi-Compton-Primakoff scattering, Gravi-bremsstrahlung in a static electric field and
nucleon-nucleon bremsstrahlung. Kaluza-Klein gravitons can couple to photons which leads to 
the first process, similarly electron-positron pair can annihilate into Kaluza-Klein gravitons,
in the third process scattering of photons by electrons may lead to Kaluza-Klein graviton 
emission via Compton or Primakoff processes (conversion of photons into Kaluza-Klein gravitons)
 - respective Feynaman diagrams can be found in 
\cite{Barger}. First of the last two processes consists in bremsstrahlung emission of 
Kaluza-Klein gravitons in static electric field of ions, the next one is similar - 
the bremsstrahlung emission of gravitons by nulceons (in electric field of nucleons). 
One can expect that first process becomes effective 
in hot stars where the photon density is large enough, the second one in stars with abundant 
electron-positron pairs (i.e. mainly the protoneutron or young neutron stars) etc.

Because white dwarfs are dense and cool one can expect that dominant process of
Kaluza-Klein graviton emission is gravi-bremsstrahlung of electrons.
The specific (mass) emissivity estimated by Barger et al. \cite{Barger} is
\begin{eqnarray} \label{epsilon}
\epsilon &=& 5.86 \; 10^{-75} \frac{T^3 n_e}{\rho M_s^4} \sum_j n_j Z_j^2 \;\;\;\;\;
\;\; for \; n=2 \\
 \epsilon &=& 9.74 \; 10^{-91} \frac{T^4 n_e}{\rho M_s^5} \sum_j n_j Z_j^2 \;\;\;\;\;
\;\; for \; n=3
\end{eqnarray}
where: $T$ is the temperature of isothermal core, $\rho$ is the density, $n_e$ and $n_j$ are 
the number densities of electrons and ions respectively. 
Total Kaluza-Klein graviton luminosity can be obtained as
\begin{equation} \label{luminosity}
L_{KK} = \int_0^{M_{WD}} \; \epsilon \; dm
\end{equation}
Extensive theoretical studies of non-radial oscillations of white dwarfs had been carried out 
long time ago \cite{Baglin69}. 
Appropriate theory of stellar oscillations consists in linearising the Poisson 
equation as well as equations 
of momentum, energy and mass conservation with respect to small non-radial perturbations. 
These perturbations (which in ZZ Ceti stars are the so called g-modes) can have oscillatory 
behavior $\propto \exp(- i \sigma t)$ where 
$\sigma^2 \propto - A$. 
The quantity $A$ (which also illustrates the connection between oscillations and convective 
instability) 
is determined by thermodynamical properties of stellar matter:
 $A = \frac{d ln \rho}{d r} - \frac{1}{\Gamma_1} \frac{d ln p}{dr}$
where: $\rho$ denotes density, $r$ - radial coordinate, $p$ is the pressure and $\Gamma_1$ 
is the adiabatic index \cite{Cox}. Now, for a zero-temperature degenerate electron gas $A = 0$ 
meaning that no g-modes are supported. However, if non-zero thermal effects are taken into account 
one can show \cite{Baglin69} that $A \propto T^2$ and consequently $\frac{1}{P} \propto T$ i.e. 
the periods scale like $1/T$ where $T$ is the core temperature. Consequently, 
the increase of the pulsation period can be calculated from the following formula:
\begin{equation}
\frac{\dot P}{P} \propto -  \frac{\dot T}{T} = -  \frac{L}{c_v M_{WD} T}
\end{equation}
where $L$ is the total luminosity, $c_v$ - average heat capacity and $M_{WD}$ is the mass of
the white dwarf. In the second part of the above formula (equality) the famous Mestel cooling 
law has been applied \cite{Mestel}.
As mentioned above we assume that anomalous cooling rate observed in the
form of pulsation period increase is due to Kaluza-Klein graviton emission. In such case we
can write \cite{Isern}
\begin{equation}\label{cooling}
\frac{L_{KK}}{L_{\gamma}} = \frac{\dot P_{obs}}{\dot P_0} - 1
\end{equation}
where $L_{\gamma}$ represents standard photon cooling, $\dot P_{obs}$ represents the
observed speed of pulsation period increase, $\dot P_0$ is an analogous quantity but
without Kaluza-Klein cooling (theoretical).

Since its discovery in 1976 \cite{McGraw} G117-B15A has been extensively studied. 
Regarding its variability the observed periods are 215.2, 271 and 304.4 s together with higher 
harmonics and linear combinations thereof \cite{Kepler82}. 
%Using the data accumulated over the period 
%1976 -- 1991 Kepler derived an upper limit to the rate of increase of main pulsation period $P=215.2\;s$ as ${\dot P} =
%(12.0 \pm 3.5) \times 10^{-15}\;\; s\;s^{-1}$. This rate appeared to be about 2 -- 3 times as big as that
%accommodated within standard CO dwarf models which predicted ${\dot P}= (2-6) \times 10^{-15}\;\;s\;s^{-1}$.   
First estimates of the rate of increase of main pulsation period $P=215.2\;s$ 
motivated Isern et al. \cite{Isern} to discuss the effect of axion emission from this star.
Very recently, with a much longer time interval of acquired data, Kepler et al. \cite{Kepler2000} recalculated the 
rate of period increase and found significantly lower value of 
${\dot P} = (2.3 \pm 1.4) \times 10^{-15} \;\;s\;s^{-1}$. 
Hence it has been claimed that the 215.2 s mode of G117-B15A is perhaps the 
most stable oscillation ever recorded in the optical band (with a stability compared to 
milisecond pulsars \cite{Corsico}).  

This circumstance makes it possible to derive a bound on string mass scale $M_s$. 
Previous upper limits on ${\dot P}$ would only allow to test whether such ``peculiar'' 
 behavior (if true) could be a manifestation of large 
extra dimensions and whether it was consistent with other astrophysical bounds.

The white dwarf pulsator G117-B15A has a mass of
$0.59 \; M_{\odot}$, effective temperature $T_{eff} = 11 620\;K$ \cite{Bergeron}
 and luminosity
$log (L/ L_{\odot}) = -2.8$ \cite{McCook} (i.e. $L_{\gamma} = 6.18 \times 10^{30}\;\;erg \;s^{-1}$).
Typical model for such CO star  
predicts the central
temperature $T = 1.2 \times 10^7\;\;K$ \cite{Corsico} and matter density $\rho = 0.97 \times 10^6 \;\;g \;cm^{-3}$.
We assume that mean molecular weight per electron is $\mu_e \approx 2$. This allows us to estimate
electron number density $n_e$. 
Following detailed calculations by Salaris \cite{Salaris} chemical composition of the white dwarf core has been taken 
as $83\%$ by mass of oxygen and remaining $17\%$ of carbon. 
Then by virtue of electric neutrality of the star one is able to estimate $n_j$.
Now we can estimate 
the power radiated away from the white dwarf core in the form of Kaluza-Klein gravitons $L_{KK}$. 
In order to derive the lower bound on $M_s$ one can take (in the role of $\dot P_{obs}$ in 
(\ref{cooling}))
an upper $2 \sigma$ limit of ${\dot P}$ equal to 
$5.1 \times 10^{-15}\;\;s\;s^{-1}$ \cite{Kepler2000}. 
Following \cite{Corsico} we assume $\dot P_0 = 3.9 \times 10^{-15}\;\;s\;s^{-1}$. 
This means that recently established secular stability of the 
fundamental oscillation mode of G117-B15A implies $L_{KK} < 0.308 \; L_{\gamma}$ which translates to:
$$
M_s > 14.3 \;\; TeV/c^2 \;\;\; for \; n=2
$$

Respective graviton emission rates for $n=3$ (and greater) theories turn out to be negligible, hence we do not
quote the resulting numbers.

Our bound has been derived 
within a very simple law relating the observed rate of the pulsational period and the 
the rate of the change of the period given by models when Kaluza-Klein gravitons 
are considered - essentially that proposed by Isern et al. \cite{Isern}. 
The most self-consistent 
approach would be to run a set of evolutionary models of white-dwarfs and compare their 
predictions concerning $\dot P$ with the observed value $\dot P_{obs}$. In the case of axion 
emission this has been done by Corsico et al. in \cite{Corsico} 
and the result obtained that way 
turned out to be in a very good agreement with 
the one obtained by applying the above mentioned much simpler method. The reason for this 
could be understood by very simple argument that the 
axion (or Kaluza-Klein gravitons as in our case) emissivity is dominated by the bremsstrahlung  
process in isothermal degenerate core.  
Based on the evolutionary white dwarf models 
the authors of \cite{Corsico} pointed out to a very important 
circumstance that additional cooling 
from the "exotic" physics affects the temperature profile in the innermost degenerate parts 
of the star. The structure of the outer (partially degenerate) 
layers depends on the temperature profile. However this profile is to great extent 
determined by the effective temperature (fixed by observations). In consequence the 
periods of oscillations (which depend on the structure of the star) remain almost 
unchanged when additional cooling (which in any case is a small correction to the energy 
budget) is present. On the other hand the value of $\dot P$ is sensitive to additional cooling. 
This justifies the assumption that one can first identify the structure of the fiducial model 
(taken by us as that cited in \cite{Corsico}) without Kaluza-Klein graviton emission 
and then incorporate 
graviton emission in discussing the rate of secular changes of the period.

\section{Conclusions}

It is interesting to compare our result with existing astrophysical constraints on the string compactification 
scale within the framework of the theory proposed by Arkani-Hamed, Dimopoulos and Dvali 
\cite{Dvali}.
Helioseismological and red-giant type considerations were performed by Cassisi et al. \cite{Cassisi}.
They calculated detailed solar models
taking into account energy loss in Kaluza-Klein gravitons explicitly in their code and obtained the lower bound for $M_s$
equal to $0.3 \;TeV/c^2$. 
More stringent limit was derived from simulating the globular cluster Hertzsprung-Russel diagram
\cite{Cassisi} (implementing Kaluza-Klein graviton emissivity into FRANEC evolutionary code).
By virtue of comparing predicted luminosity of RGB tip with observations
Cassisi et al. obtained a "red-giant" bound for $M_s$ to be $3-4\;\;TeV/c^2$. Our estimate of the string
energy scale implied by recently reported stability of ZZ Ceti star G117-B15A equal to $14.3 \;\; TeV/c^2$ is much 
stronger than above mentioned stellar evolutionary bounds.  

Among existing astrophysical bounds on Kaluza-Klein theories with large extra dimensions
only the supernova constraints are more restrictive. They demand $M_s > 30 - 130 \;
\; TeV/c^2$ \cite{Barger, Cullen} and are based on a different mechanism of Kaluza-Klein graviton emission -- the 
nucleon-nucleon bremsstrahlung. 
On the other hand the lesson learned in testing the physics of axions shown that
first straightforward supernova bounds were reduced by an order of magnitude when more accurate 
nuclear physics was employed
\cite{Raffelt97}. Quite recently, the paper with improved calculations of nucleon-nucleon bremsstrahlung appeared 
\cite{Savage} suggesting that former SN1987a bounds should be lowered by a factor of $0.65.$ It should also be noted 
that cosmological considerations \cite{Fairbairn} 
apparently provide much more stringent bounds than the supernova SN1987a.

The bound derived in this paper is based on white dwarf cooling. The physics underlying this process is very simple 
hence one can expect that the result is robust 
(within the framework of the theory proposed by Arkani-Hamed, Dimopoulos and Dvali \cite{Dvali}). 
Kaluza-Klein graviton emissivity at the relevant densities and 
temperatures of white dwarfs is dominated by the gravi-bremsstrahlung process taking place in the degenerate and 
isothermal core. The mass of the core (essentially equal to the mass of the star, since the outer helium layer comprises 
less than $0.01\;M_{WD}$, its temperature and chemical composition can be reliably estimated by fitting 
evolutionary models to observational characteristics such like effective temperature or oscillation periods.    
The cooling rate of B117-G15A pulsating star 
is constrained by measurements which are performed with great accuracy. The most recent determination of the secular rate 
of change of the period \cite{Kepler2000} took into account all the periodicities and the error bars reported therein 
can be considered as safe. It can be argued \cite{Corsico} that remaining uncertainty -- mostly from mode identification 
procedure and the precise physical characteristics (mass or temperature) is of order of $1 \times 10^{-11}\;\;s\;s^{-1}$. 
Other 
effects like the contribution of the proper motion \cite{Pajdosz, Kepler2000} or the rate of reaction ${}^{12}C
(\alpha,\gamma){}^{16}O$ which determines the stratification of the core at final stage of the asymptotic giant branch, 
both contribute an order of magnitude smaller value to the final uncertainty. 

The detailed discussion of using a pulsational code coupled to evolutionary code aimed at constraining an axion mass 
by observed stability of the fundamental mode of G117-B15A can be found in a recent paper by C{\'o}rsico et al. 
\cite{Corsico}. 
One of the conclusions formulated in \cite{Corsico} was that the bounds on axion mass derived from simple estimates 
like performed in the present paper (in a different context) or in \cite{Isern} are in good agreement with evolutionary 
calculations. Although axion emissivity has different temperature dependence than that of Kaluza-Klein gravitons, one can 
expect the same for gravitons. Hence the B117-G15A pulsator remains an important tool for testing fundamental physics.  

\section{Acknowledgements}
The authors would like to thank Prof. Pawe{\l} Moskalik for useful discussions and comments.


\begin{thebibliography}{X}

\bibitem{Green87}
M.B. Green, J.H. Schwarz and E. Witten, Superstring Theory, Cambridge
University Press, 1987.

\bibitem{Antoniadis}
I. Antoniadis, Phys.Lett. {\bf B246}, 377, 1990.

\bibitem{Witten}
E. Witten,Nucl.Phys. {\bf B471}, 135, 1996.

\bibitem{Dvali}
N. Arkani-Hamed, S. Dimopoulos and G. Dvali, Phys. Lett. {\bf B429}, 263, 1998;\\
I. Antoniadis, N. Arkani-Hamed, S. Dimopoulos and G. Dvali, Phys. Lett. {\bf B436}, 257, 1998.

\bibitem{LEP}
G.F. Giudice, Int.J.Mod.Phys. {\bf A15S1},  440-463, 2000

\bibitem{Raffelt}
G.G. Raffelt, Particle Physics from the Stars, Annu.Rev.Nucl.Part.Sci. {\bf 49}, 163-216, 1999

\bibitem{helioseismology}
S. Degl'Innocenti, W. Dziembowski, G.Fiorentini and B.Ricci, Astr.Phys. {\bf 7}, 77, 1997

\bibitem{Isern}
J. Isern, M. Hernandez, E. Garcia-Berro, ApJ {\bf 392}, L23-25, 1992

\bibitem{Baglin69}
A. Baglin and J. Hayvaerts, Nature {\bf 222}, 1258, 1969

\bibitem{Winget83}
D.E. Winget, C.J. Hansen and H.M. Van Horn, Nature {\bf 303}, 781, 1983

\bibitem{Barger}
V.Barger, T.Han, C.Kao and R.J.Zhang, Phys.Lett. {\bf B 461}, 34-42, 1999;  hep-ph/9905474

\bibitem{Cullen}
S. Cullen, M. Perelstein, Phys.Rev.Lett. {\bf 83}, 268, 1989

\bibitem{Savage}
Ch. Hanhart, D.R. Phillips, S. Reddy, M.J. Savage, 
%"Extra dimensions, SN1987a, and nucleon-nucleon scattering data", nucl-th/0007016
Nucl.Phys. {\bf B595}, 335-359, 2001; nucl-th/0007016

\bibitem{Cassisi}
S.Cassisi, V.Castellani, S.Degl'Innocenti, G.Fiorentini and B.Ricci, Phys.Lett. {\bf B 481}, 323-332, 2000; astro-ph/0002182

\bibitem{Giudice}
G.F. Giudice, R. Rattazzi and J.D. Wells, Nucl. Phys. {\bf B544}, 3, 1999\\
T. Han, J.D.Lykken and R.-J. Zhang, Phys. Rev. {\bf D59}, 105006, 1999.

\bibitem{Cox}
J.P. Cox Theory of Stellar Pulsations (Princeton University Press) 1980

\bibitem{Mestel}
L. Mestel, MNRAS, {\bf 112}, 583, 1952

\bibitem{McGraw}
J.T. McGraw, E.L. Robinson, ApJ, {\bf 205}, L155, 1976

\bibitem{Kepler82}
S.O. Kepler, E.L. Robinson, R.E. Nather, J.T. McGraw, ApJ, {\bf 254}, 676, 1982

\bibitem{Kepler91}
S.O. Kepler, et al. ApJ, {\bf 378}, L45, 1991

\bibitem{Kepler2000}
S.O. Kepler, A. Mukadam, D.E. Winget, R.E. Nather, T.S. Metcalfe, M.D. Reed, S.D. Kawaler, P.A. Bradley, 
ApJ, {\bf 534}, L185, 2000

\bibitem{Corsico} 
A.H. C{\'o}rsico, O.G. Benvenuto, L.G. Althaus, J. Isern, E. Garcia-Berro, 
%%%"The potential of the variable DA white dwarf G117-B15A as a tool for Fundamental Physics", 2001, 
New Astron. {\bf 6}, 197-213, 2001 

\bibitem{Bergeron}
P. Bergeron, F. Wesemael, R. Lamontagne, G. Fontaine, R.A. Saffer, N.F. Allard, ApJ, {\bf 449}, 258, 1995

\bibitem{McCook}
G.P. McCook, E.M. Sion, ApJS, {\bf 121}, 1, 1999

\bibitem{Salaris}
M. Salaris, I. Dominguez, E. Garcia-Berro, M. Hernanz, J. Isern, R. Moschkovitz, ApJ, {\bf 486}, 413, 1997

\bibitem{Pajdosz}
G. Pajdosz, Astron.Astrophys., {\bf 295}, L17, 1995

%\bibitem{Fontaine}
%G. Fontaine, Brassard P., Wesemael F, Kepler S.O., Wood M.A., in White Dwarfs, ed. G.Vauclair
%and E.Sion (Dordrecht:Kluwer)5, 1991

\bibitem{Raffelt97}
G.G. Raffelt, in "Beyond the Desert" Proc.of the Conference, Ringberg Castle, Tegernsee, Germany June 8-14, 1997;
astro-ph/9707268v2

\bibitem{Fairbairn}
M. Fairbairn, 
%"Cosmological constraints on large extra dimensions", hep-ph/0101131
Phys.Lett. {\bf B508}, 335-339, 2001



\end{thebibliography}
\end{document}